# Rates of convergence of nonextensive statistical distributions to Lévy distributions in full and half spaces

Sumiyoshi Abe[†] and A. K. Rajagopal[‡]

[†]College of Science and Technology, Nihon University,

Funabashi, Chiba 274-8501, Japan

[‡]Naval Research Laboratory, Washington, DC 20375-5320, USA

**Abstract.** The Lévy-type distributions are derived using the principle of maximum Tsallis nonextensive entropy both in the full and half spaces. The rates of convergence to the exact Lévy stable distributions are determined by taking the $N$-fold convolutions of these distributions. The marked difference between the problems in the full and half spaces is elucidated analytically. It is found that the rates of convergence depend on the ranges of the Lévy indices. An important result emerging from the present analysis is deduced if interpreted in terms of random walks, implying the dependence of the asymptotic long-time behaviors of the walks on the ranges of the Lévy indices if $N$ is identified with the total time of the walks.

## 1. Introduction

There is growing interest in Lévy-type random walks and anomalous diffusion phenomena. They are often encountered in many physical systems, including polymer-like breakable micelles dissolved in salted water [1], tracer particles in rotating flow [2], subrecoil laser cooling of atoms [3], two-particle dispersion in fully developed turbulence [4], and single molecule line shape cumulants in glasses [5]. (See also Refs. [6-8].) To describe and understand these phenomena physically and mathematically, it seems necessary to generalize the framework of traditional statistical mechanics. For example, it was pointed out [9-14] that introduction of Riemann-Liouville fractional calculus [15] to the diffusion and Fokker-Planck equations can lead to solutions of the Lévy-type distributions (though not the exact Lévy distributions). Another organizing principle to understand such phenomena was based on the method of maximizing the Boltzmann-Shannon entropy for extensive systems. It was found in Ref. [16] that, in order to obtain Lévy distributions using this method, one has to impose a constraint which does not admit a natural physical interpretation. The main reason for this unsatisfactory state of affairs is due to the divergent second moment of Lévy-type distribution. Recently, Tsallis [17] introduced an alternate to the Boltzmann-Shannon entropy to treat nonextensive systems [18,19]. It was soon realized [20-24] that this form of the entropy can satisfactorily incorporate situations where divergent first few moments of the distribution occur. In this approach, a power-law distribution is derived as the maximum Tsallis entropy state. It has the same asymptotic behavior as the exact Lévy distribution for large values of the relevant random variable, and therefore its lowest moments diverge. In the language of random walks, this optimal distribution may describe a single jump, and thus, after $N$-jumps (i.e., $N$-fold convolution), it is expected to converge to the exact Lévy stable distributions in the large-$N$ limit. Considering $N$ as the total time of the walks, such a limit implies the long-time behaviors of the walks. The idea behind this is the Lévy-Gnedenko generalized central limit theorem [25,26], which states that by $N$-fold convolution a distribution with divergent lowest moments tends to one of the Lévy stable class of distributions in the limit $N \to \infty$ *if such a limit is convergent*. However, the convergence has not been discussed in the previous works [20-24].

Now, associated with the convergence property, there is also an important issue regarding *the rate of convergence*. This is actually a generic physical problem related to the limiting theorems in probability theory. For example, in Ref. [27], a truncated Lévy distribution, i.e. a Lévy distribution whose long tail cut off, was studied numerically. Since such a distribution has a finite second moment, the relevant mathematics there is the ordinary central limit theorem, that is, convergence of the distribution to a Gaussian distribution by many-fold convolution. The authors of Ref. [27] discovered an intersting fact that convergence of the truncated Lévy distribution to a Gaussian distribution is extraordinarily slow. This result tells us that there is an essential difference between Lévy-type and non-Lévy-type distributions in a finite space.

The purpose of the present paper is two-fold: First, to study the rate of convergence of the maximum Tsallis entropy distribution to the exact Lévy distribution. In particular, we consider this problem in full space $(-\infty, \infty)$, which was not addressed in Refs. [20-24]. Second, to examine the derivation of the exact Lévy distribution in half space $(0, \infty)$ by this method, which has not been discussed in the literature. The half-space problem is very different from the full-space one and has important physical implications. Specifically, the half space analysis is required for Hamiltonian systems whose spectra are bounded from below. The rate of convergence is also clarified in the half-space case. We analytically determine the corrections to the exact Lévy stable distributions in terms of the number of times of convolution. We shall show that, quite remarkably, the rates of convergence depend the ranges of the Lévy indices in the cases of both the full and half spaces.

The paper is organized as follows. In Sec. 2, the rate of convergence of the maximum Tsallis entropy distribution to the exact Lévy distribution is studied in the full space. Then, in Sec. 3, we formulate the Lévy problem in the framework of the principle of maximum Tsallis entropy in the half space as well as determine the rate of convergence. Section 4 is devoted to conclusions.

## 2. Maximum Tsallis entropy distribution in full space

A complete derivation of the maximum Tsallis entropy distribution in the full space $(-\infty, \infty)$ within the framework of the *normalized q-expectation value formalism* [28] was given in Ref. [24]. The authors of Ref. [24] considered the optimization of the Tsallis entropy

$$S_q[p] = \frac{1}{1-q}\left\{\int_{-\infty}^{\infty} \frac{dx}{\sigma}[\sigma p(x)]^q - 1\right\}, \tag{1}$$

under the constraints on the normalization condition

$$\int_{-\infty}^{\infty} dx\, p(x) = 1, \tag{2}$$

and on the generalized second moment defined in terms of the normalized $q$-expectation value

$$\int_{-\infty}^{\infty} dx\, x^2 P_q(x) = \sigma^2, \tag{3}$$

where $\sigma$ is a positive constant giving a length scale and $P_q(x)$ is the escort distribution [29] associated with $p(x)$, i.e.,

$$P_q(x) = \frac{[p(x)]^q}{\int_{-\infty}^{\infty} dx' [p(x')]^q}. \tag{4}$$

$q$ in eqs. (1) and (4) is the entropic index, which is assumed to be

$$\frac{5}{3} < q < 3 \tag{5}$$

for the problem of a Lévy-type distribution. The maximum Tsallis entropy distribution was then found to be [24]

$$p(x) = \frac{1}{\sigma}\sqrt{\frac{q-1}{\pi(3-q)}} \frac{\Gamma\left(\frac{1}{q-1}\right)}{\Gamma\left(\frac{3-q}{2q-2}\right)} \frac{1}{\left(1 + \frac{q-1}{3-q}\frac{x^2}{\sigma^2}\right)^{1/(q-1)}}, \tag{6}$$

where $\Gamma(z)$ is the Euler gamma function. It is noted that this distribution has the divergent ordinary second moment, $\int_{-\infty}^{\infty} dx\, x^2 p(x) = \infty$, but the generalized second moment in eq. (3) is finite. For large values of $x$, the above distribution asymptotically behaves as

$$p(x) \sim x^{-2/(q-1)}. \tag{7}$$

On the other hand, the exact symmetric Lévy distribution in the full space [25,26]

$$L_\gamma(x) = \frac{1}{2\pi} \int_{-\infty}^{\infty} dk \, \exp(-ikx) \exp(-a|k|^\gamma) \tag{8}$$

has the following asymptotic form:

$$L_\gamma(x) \sim x^{-1-\gamma}, \tag{9}$$

where $a$ is a positive constant and $\gamma$ the Lévy index satisfying

$$0 < \gamma < 2. \tag{10}$$

Let us examine the convergence properties of the distribution in eq. (6) to the exact Lévy distribution $L_\gamma(x)$ in view of the generalized central limit theorem [25,26]. First of all, comparing eq. (7) with eq. (9), $q$ is expected to be related to the Lévy index as follows:

$$q = \frac{3+\gamma}{1+\gamma}. \tag{11}$$

In order to apply the theorem, it is necessary to consider the sum of $N$ scaled random variables $\{X_i\}_{i=1,2,\text{L},N}$, that is,

$$X = \frac{X_1 + X_2 + \text{L} + X_N}{B_N}. \tag{12}$$

$X_i (i = 1, 2, \text{L}, N)$ and $X$ respectively take the values $x_i$ and $x$, which run over whole of real numbers. $X_i$'s are assumed to be independently and identically distributed. The distribution of $X$ is essentially given by $N$-fold convolution. The scale factor $B_N$ has to be chosen in such a way that the limit distribution is independent of the number $N$ of convolutions. Suppose each $X_i$ obeys the Lévy distribution in eq. (8). Then, the distribution of $X$ is given by

$$L_\gamma^{(N)}(x) = B_N \left(L_\gamma * \text{L} * L_\gamma\right)(B_N x), \tag{13}$$

where the convolution in the full space reads

$$(f * g)(x) = \int_{-\infty}^{\infty} dx' \, f(x - x') \, g(x'). \tag{14}$$

To see the above-mentioned independence, we employ the standard method of characteristic functions. The characteristic function of $L_\gamma^{(N)}(x)$ is calculated to be

$$\chi_L^{(N)}(k) = \int_{-\infty}^{\infty} dx \, \exp(ikx) \, L_\gamma^{(N)}(x)$$

$$= \left[\chi_L\left(\frac{k}{B_N}\right)\right]^N, \tag{15}$$

where $\chi_L(k)$ is the characteristic function of $L_\gamma(x)$:

$$\chi_L(k) = \exp\left(-a|k|^\gamma\right). \tag{16}$$

Therefore, we have

$$B_N = N^{1/\gamma}, \tag{17}$$

where, without loss of generality, the constant of proportionality is set equal to unity. A set of limit distributions $\{L_\gamma(x): 0 < \gamma < 2\}$ forms a stable class.

To examine the convergence properties of the distribution in eq. (16) by many-time self-convolutions, again we calculate its characteristic function. After some manipulations, we find

$$\chi(k) = \int_{-\infty}^{\infty} dx \, \exp(ikx) \, p(x)$$

$$= 2^{1-\nu} \frac{(\lambda |k|)^{\nu}}{\Gamma(\nu)} K_{\nu}(\lambda |k|), \tag{18}$$

where $K_{\nu}(z)$ is the modified Bessel function [30] and

$$\nu = \frac{3-q}{2q-2} = \frac{\gamma}{2}, \tag{19}$$

$$\lambda = \sigma\sqrt{2\nu}. \tag{20}$$

Clearly, the range of $\nu$ is

$$0 < \nu < 1. \tag{21}$$

Next, let us discuss the limit $N \to \infty$ of the quantity

$$f(k; N) \equiv \left[\chi\left(\frac{k}{N^{1/\gamma}}\right)\right]^{N}. \tag{22}$$

For this purpose, it is convenient to take its logarithm, that is,

$$\ln f(k; N) = N \ln \left[ 2^{1-\gamma/2} \frac{\left(\frac{\lambda |k|}{N^{1/\gamma}}\right)^{\gamma/2}}{\Gamma\left(\frac{\gamma}{2}\right)} K_{\gamma/2}\left(\frac{\lambda |k|}{N^{1/\gamma}}\right) \right]. \tag{23}$$

Using the formula [30]

$$K_{\nu}(z) = \frac{\pi}{2} \frac{I_{-\nu}(z) - I_{\nu}(z)}{\sin(\nu\pi)}, \tag{24}$$

where $I_\nu(z)$ is the Bessel function of imaginary argument, and evaluating the series expansion [30]

$$I_\nu(z) = \sum_{n=0}^{\infty} \frac{\left(\frac{z}{2}\right)^{2n+\nu}}{n!\,\Gamma(n+\nu+1)}, \qquad (25)$$

we have

$$\ln f(k;N) = N \ln\left[1 - \frac{1}{N}\frac{\Gamma\left(1-\frac{\gamma}{2}\right)}{\Gamma\left(1+\frac{\gamma}{2}\right)}\left(\frac{\lambda|k|}{2}\right)^\gamma + \frac{2}{2-\gamma}\frac{1}{N^{2/\gamma}}\left(\frac{\lambda|k|}{2}\right)^2 + \mathrm{L}\right]$$

$$= -\frac{\Gamma\left(1-\frac{\gamma}{2}\right)}{\Gamma\left(1+\frac{\gamma}{2}\right)}\left(\frac{\lambda|k|}{2}\right)^\gamma + \frac{2}{2-\gamma}\frac{1}{N^{2/\gamma-1}}\left(\frac{\lambda|k|}{2}\right)^2$$

$$-\frac{1}{2N}\left[\frac{\Gamma\left(1-\frac{\gamma}{2}\right)}{\Gamma\left(1+\frac{\gamma}{2}\right)}\right]^2 \left(\frac{\lambda|k|}{2}\right)^{2\gamma} + \mathrm{L}\,. \qquad (26)$$

Thus, we find that, in the limit $N \to \infty$, the maximum Tsallis entropy distribution in fact converges to the exact symmetric Lévy distribution with

$$a = \frac{\Gamma\left(1-\frac{\gamma}{2}\right)}{\Gamma\left(1+\frac{\gamma}{2}\right)}\left(\frac{\lambda}{2}\right)^\gamma. \qquad (27)$$

Now, from the above analysis, we also find the rate of convergence. From eq. (26), it follows that the decay of the correction to the exact Lévy distribution in the large-$N$ limit behaves in two different ways depending on the range of the Lévy index:

$$O(N^{-1}) \quad (0 < \gamma < 1), \tag{28}$$

$$O(N^{1-2/\gamma}) \quad (1 < \gamma < 2). \tag{29}$$

Therefore, we conclude that the convergence is slower for $1 < \gamma < 2$.

### 3. Maximum Tsallis entropy distribution in half space

It is known [26] that the properties of probability distributions defined in the half space $(0, \infty)$ are quite different from those in the full space $(-\infty, \infty)$. Physically, such distributions are important in cases where the basic random variables describing systems are bounded from below. A typical example is the Hamiltonian of a stable system. So far, the maximum Tsallis entropy approach to study of Lévy-type distributions has not been discussed in the half space. In this section, we provide such a demonstration in parallel with the discussion in the previous section. We shall use for the half-space problem the same notation as for the full space, but it should not cause any confusion.

Analogously to eq. (1), the Tsallis entropy in the half space is given by

$$S_q[p] = \frac{1}{1-q} \left\{ \int_0^\infty \frac{dx}{\sigma} [\sigma p(x)]^q - 1 \right\}. \tag{30}$$

We maximize this quantity under the constraints on the normalization condition

$$\int_0^\infty dx\, p(x) = 1, \tag{31}$$

and on the generalized first moment defined in terms of the normalized $q$-expectation value

$$\langle X \rangle_q = \int_0^\infty dx\, x\, P_q(x) \equiv \sigma, \tag{32}$$

where $X$ is the basic random variable which takes a positive real value $x$. Here, $P_q(x)$ is the escort distribution in the half space

$$P_q(x) = \frac{[p(x)]^q}{\int_0^\infty dx'\, [p(x')]^q}. \tag{33}$$

The resulting distribution is found to be

$$p(x) = \frac{1}{Z_q(\beta)}\left[1 - (1-q)\left(\tilde{\beta}/c_q\right)\left(x - \langle X \rangle_q\right)\right]^{1/(1-q)}, \tag{34}$$

where $Z_q(\beta)$ is the normalization constant, $\tilde{\beta} = \beta/\sigma^{q-1}$ with $\beta$ the Lagrange multiplier associated with the constraint in eq. (32) and

$$c_q \equiv \int_0^\infty dx\, [p(x)]^q. \tag{35}$$

A long-tailed distribution is realized for $q > 1$. The normalizability condition requires $q < 2$. In addition, the divergence of the ordinary first moment, $\int_0^\infty dx\, x\, p(x) = \infty$, places another condition $q > 3/2$. Therefore, the range of interest is

$$\frac{3}{2} < q < 2. \tag{36}$$

$c_q$ and $\langle X \rangle_q$ are the quantities, which have to be calculated by using $p(x)$. In what follows, we determine these quantities self-consistently. The normalization condition on $p(x)$ leads to the identical relation

$$c_q = \left[ Z_q(\beta) \right]^{1-q}. \tag{37}$$

Let us rewrite eq. (34) in the form

$$p(x) = \frac{1}{Z_q(\beta)} \left[ \frac{c_q}{(q-1)\tilde{\beta}} \right]^{1/(q-1)} \frac{1}{(\xi + x)^{1/(q-1)}}, \tag{38}$$

where

$$\xi = \frac{c_q}{(q-1)\tilde{\beta}} - \langle X \rangle_q. \tag{39}$$

From eqs. (38) and (40), consistency requires that $\xi$ be positive. This point will be ascertained subsequently. Using the distribution in eq. (38), we have the following result:

$$\langle X \rangle_q = \frac{c_q}{\tilde{\beta}}. \tag{40}$$

This is a relation of crucial importance for obtaining the self-consistent solution. From it, it is clear that $\xi$ in eq. (39) is in fact positive, as promised above. Thus, the maximum Tsallis entropy distribution in the half space is written in the form

$$p(x) = \frac{2-q}{q-1} \xi^{(2-q)/(q-1)} \frac{1}{(\xi+x)^{1/(q-1)}}. \tag{41}$$

The distribution of this type is often referred to as the Zipf-Mandelbrot distribution in the literature. Substituting eq. (41) into eq. (32), we have

$$\xi = \frac{2-q}{q-1} \sigma. \tag{42}$$

Consequently, we obtain the following distribution:

$$p(x) = \frac{1/\sigma}{\left(1 + \frac{q-1}{2-q} \frac{x}{\sigma}\right)^{1/(q-1)}}. \tag{43}$$

Next, let us recall the exact Lévy distribution $L_\gamma(x)$ in the half space [26]. It is defined as follows:

$$L_\gamma(x) = \frac{1}{2\pi} \int_{-\infty}^{\infty} dk \, \exp(-ikx) \, \chi_L(k), \tag{44}$$

$$\chi_L(k) = \exp\left\{-a|k|^\gamma \exp\left[i\varepsilon(k)\frac{\theta\pi}{2}\right]\right\}, \tag{45}$$

where $a$ is a positive constant, $\gamma$ the Lévy index, $\theta$ a constant satisfying $|\theta| \leq \gamma$ and $\varepsilon(k) = k/|k|$ the sign function of $k$. In contrast to the case of the full space, the range of the Lévy index in the half space is

$$0 < \gamma < 1. \tag{46}$$

This distribution has the following asymptotic behavior for large values of $x$:

$$L_\gamma(x) \sim x^{-1-\gamma}, \tag{47}$$

and therefore the ordinary first moment is divergent: $\int_0^\infty dx\, x\, L_\gamma(x) = \infty$.

Comparing eq. (47) with the asymptotic form of eq. (43), we expect the following relation between $q$ and the Lévy index:

$$q = \frac{2+\gamma}{1+\gamma}. \tag{48}$$

As in the case of the full space, $\{L_\gamma(x): 0 < \gamma < 1\}$ forms a stable class. To see this, we again consider the scaled sum of independently and identically distributed $N$ positive random variables $\{X_i > 0\}_{i=1,2,\mathrm{L},N}$, that is,

$$X = \frac{X_1 + X_2 + \mathrm{L}\ X_N}{B_N}. \tag{49}$$

Each $X_i$ $(i = 1, 2, \mathrm{L}, N)$ is assumed to obey the exact Lévy distribution $L_\gamma(x)$ in eq. (44). The scaling factor $B_N$ is chosen in such a way that the limit distribution is independent of the number $N$ of convolutions. The distribution of $X$ is given by $N$-fold convolution of $L_\gamma(x)$:

$$L_\gamma^{(N)}(x) = B_N \big(L_\gamma * \mathrm{L}\ * L_\gamma\big)(B_N x), \tag{50}$$

where the convolution here is defined by

$$(f * g)(x) = \int_0^x dx'\, f(x-x') g(x'). \tag{51}$$

The characteristic function of $L_\gamma^{(N)}(x)$ is calculated to be

$$\chi_L^{(N)}(k) = \int_0^\infty dx\, \exp(ikx)\, L_\gamma^{(N)}(x)$$

$$= \left[\chi_L\left(\frac{k}{B_N}\right)\right]^N. \tag{52}$$

(In calculating the inverse Fourier transformation, note that $L_\gamma(x) = 0$ for $x < 0$.) Thus, we find that the exact Lévy distribution is invariant under $N$-fold convolution if

$$B_N = N^{1/\gamma}. \tag{53}$$

Now, we are in a position to discuss how the maximum Tsallis entropy distribution $p(x)$ in eq. (43) converges to the exact Lévy distribution in eq. (44) in accordance with the generalized central limit theorem. For this purpose, we again consider the characteristic function of $p(x)$, which is calculated as follows:

$$\chi(k) = \int_0^\infty dx\, \exp(ikx)\, p(x)$$

$$= \int_0^\infty \frac{dx}{\sigma} \frac{\exp(ikx)}{\left(1 + \frac{q-1}{2-q}\frac{x}{\sigma}\right)^{1/(q-1)}}$$

$$= M\left(1, 2 - \frac{1}{q-1}; -ik\sigma \times \frac{2-q}{q-1}\right)$$

$$- \Gamma\left(2 - \frac{1}{q-1}\right)\left(-ik\sigma \times \frac{2-q}{q-1}\right)^{(2-q)/(q-1)} \exp\left(-ik\sigma \times \frac{2-q}{q-1}\right), \tag{54}$$

where $M(a, b; z)$ is the Kummer function [31] with the expansion

$$M(a, b; z) = 1 + \frac{a}{b}z + \frac{a(a+1)}{b(b+1)}\frac{z^2}{2!} + \text{L} . \tag{55}$$

The quantity to be considered is

$$f(k; N) \equiv \left[\chi\left(\frac{k}{N^{1/\gamma}}\right)\right]^N . \tag{56}$$

Taking the logarithms of both sides of this equation, we have

$$\ln f(k; N) = N \ln\left[M\left(1, 1-\gamma; -\frac{i\gamma k\sigma}{N^{1/\gamma}}\right)\right.$$

$$\left. - \frac{1}{N}\Gamma(1-\gamma)(-i\gamma k\sigma)^\gamma \exp\left(-\frac{i\gamma k\sigma}{N^{1/\gamma}}\right)\right], \tag{57}$$

where we have used eq. (48). Performing the expansion for large $N$, we find

$$\ln f(k; N) = -(\gamma\sigma)^\gamma \Gamma(1-\gamma)|k|^\gamma \exp\left[-i\varepsilon(k)\frac{\gamma\pi}{2}\right]$$

$$-\frac{1}{N^{-1+1/\gamma}}\frac{i\gamma k\sigma}{1-\gamma} - \frac{1}{2N}\left[\Gamma(1-\gamma)(-i\gamma k\sigma)^\gamma\right]^2 + \text{L} . \tag{58}$$

Therefore, from eqs. (45) and (58), we may identify

$$a = (\gamma\sigma)^\gamma \Gamma(1-\gamma), \tag{59}$$

$$\theta = -\gamma. \tag{60}$$

The rate of convergence to the exact Lévy distribution for large $N$ can be determined from eq. (58). We find the behavior analogous to the case of the full space in Sec. 2:

$$O(N^{-1}) \quad (0 < \gamma < 1/2), \tag{61}$$

$$O(N^{1-1/\gamma}) \quad (1/2 < \gamma < 1). \tag{62}$$

Thus, we conclude that the convergence is slower for $1/2 < \gamma < 1$.

## 4. Conclusions

We have derived the Lévy-type distributions, their convergence to the exact Lévy stable distributions and its rates both in the full and half spaces based on the principle of maximum Tsallis nonextensive entropy. This has been done by considering the $N$-fold convolutions of the distributions arising from the use of the above principle in conformity with the Lévy-Gnedenko generalized central limit theorem. A new result emerging from this analysis is that the rates of convergence depend on the ranges of the respective Lévy indices. This result interpreted in terms of random walks implies that asymptotic long-time behaviors of the (Tsallis) walks depend on the ranges of the Lévy indices if $N$ is regarded as the total time of the walks. We have also elucidated the marked difference between the problems in the full and half spaces analytically.

**Acknowledgments**

One of us (S.A.) was supported in part by the Grant-in-Aid for Scientific Research of Japan Society for the Promotion of Science and by the GAKUJUTSU-SHO Program of College of Science and Technology, Nihon University. The other (A. K. R.) acknowledges the partial support from the US Office of Naval Research. He also acknowledges the support from the International Academic Exchange Program of College of Science and Technology, Nihon University, which has enabled this collaboration. We would like to thank the referee for perceptive comments.

**References**

[1] Ott A, Bouchaud J P, Langevin D and Urbach W 1990 *Phys. Rev. Lett.* **65** 2201

[2] Solomon T H, Weeks E R and Swinney H L 1993 *Phys. Rev. Lett.* **71** 3975

[3] Bardou F, Bouchaud J P, Emile O, Aspect A and Cohen-Tannoudji C 1994 *Phys. Rev. Lett.* **72** 203

[4] Sokolov I M, Blumen A and Klafter J 1999 *Europhys. Lett.* **47** 152

[5] Barkai E, Silbey R and Zumofen G 2000 *Phys. Rev. Lett.* **84** 5339

[6] Montroll E W and Shlesinger M F 1984 in *Nonequilibrium Phenomena II: From Stochastics to Hydrodynamics* ed Lebowitz J L and Montroll E W (Amsterdam: Elsevier)

[7] Bouchaud J P and Georges A 1990 *Phys. Rep.* **195** 127

[8] *Lévy Flights and Related Topics in Physics* 1995 ed Shlesinger M F, Zaslavsky G M and Frisch U (Berlin: Springer-Verlag)

[9] Nonnenmacher T F 1990 *J. Phys.* A **23** L697

[10] Zaslavsky G M 1994 *Physica* D **76** 110

[11] Chaves A S 1998 *Phys. Lett.* A **239** 13

[12] Metzler R, Klafter J and Sokolov I M 1998 *Phys. Rev.* E **58** 1621

[13] Grigolini P, Rocco A and West B J 1999 *Phys. Rev.* E **59** 2603

[14] Barkai E, Mezler R and Klafter J 2000 *Phys. Rev.* E **61** 132

[15] Hilfer R (ed) 2000 *Applications of Fractional Calculus in Physics* (Singapore: World Scientific)

[16] Montroll E W and Shlesinger M F 1983 *J. Stat. Phys.* **32** 209

[17] Tsallis C 1988 *J. Stat. Phys.* **52** 479

[18] Salinas S R A and Tsallis C (ed) 1999 *Braz. J. Phys.* **29** Special Issue, which can be obtained from http://sbf.if.usp.br/WWW_pages/Journals/BJP/Vol29/Num1/index.htm

[19] Abe S and Okamoto Y (ed) *Nonextensive Statistical Mechanics and Its Applications* (Heidelberg: Springer-Verlag) to appear

[20] Alemany P A and Zanette D H 1994 *Phys. Rev.* E **49** R956

[21] Zanette D H and Alemany P A 1995 *Phys. Rev. Lett.* **75** 366
Cáceres M O and Budde C E 1996 *Phys. Rev. Lett.* **77** 2589
Zanette D H and Alemany P A 1996 *Phys. Rev. Lett.* **77** 2590

[22] Tsallis C, Levy S V F, Souza A M C and Maynard R 1995 *Phys. Rev. Lett.* **75** 3589

[23] Buiatti M, Grigolini P and Montagnini A 1999 *Phys. Rev. Lett.* **82** 3383

[24] Prato D and Tsallis C 1999 *Phys. Rev.* E **60** 2398

[25] Gnedenko B V and Kolmogorov A N 1968 *Limit Distributions for Sums of Independent Random Variables* (Reading, Massachusetts: Addison-Wesley)

[26] Feller W 1971 *An Introduction to Probability Theory and Its Applications Vol. II* (New York: Wiley)

[27] Mantegna R N and Stanley H E 1994 *Phys. Rev. Lett.* **73** 2946

[28] Tsallis C, Mendes R S and Plastino A R 1998 *Physica* A **261** 534

[29] Beck C and Schlögl F 1993 *Thermodynamics of Chaotic Systems: An Introduction* (Cambridge: Cambridge University Press)

[30] Gradshteyn I S and Ryzhik I M 1980 *Table of Integrals, Series, and Products* (New York: Academic Press)

[31] Abramowitz M and Stegun I A (ed) 1972 *Handbook of Mathematical Functions* (New York: Dover)